\documentclass[11pt]{article}
\usepackage[english]{babel}
\usepackage[utf8]{inputenc}
\usepackage[nottoc]{tocbibind}
\usepackage{graphicx}
\usepackage{amsmath}
\usepackage{amssymb}
\usepackage{amsxtra}

\def\institute#1{\gdef\@institute{#1}}

\title{Quantum-mechanical numerical model of interaction between dark atom and nucleus of substance}
\author{T.E. Bikbaev$^{1,*}$, M.Yu. Khlopov$^{2}$, A.G. Mayorov$^{1}$\\
$^{1}$ National Research Nuclear University MEPhI \\115409 Moscow, Russia;\\
$^{2}$ Virtual Institute of Astroparticle physics,\\ Paris 75018, France; e-mail: khlopov@apc.in2p3.fr (M.K.); \\ 
$^{*}$ Correspondence: bikbaev.98@bk.ru}

\date{November 2024}

\begin{document}
\maketitle

\begin{abstract}
The hypothesis of composite $X$He dark atoms may provide solution to the long-standing problem of direct searches for dark matter particles. The main problem of the $X$He dark atom is its ability to strongly interact with the nucleus of substance, arising from the unshielded nuclear attraction between the helium nucleus and the nucleus of matter. It is assumed that in order to prevent the destruction of the bound structure of dark atom, the effective potential of interaction between $X$He and the nucleus of substance must have dipole Coulomb barrier that prevents the fusion of dark matter atom particles with the nucleus of substance. The problem in describing the interaction between dark atom and substance nucleus is the three-body problem, for which an exact analytical solution is not available. Consequently, to assess the physical meaning of the proposed scenario, it is essential to develop a numerical approach. Our approach involves consistently developing an accurate quantum mechanical description of this three-body system, comprising bound dark atom and the external nucleus of substance. We incorporate the necessary effects and interactions to enhance the precision of the results, which helps to elucidate the most significant aspects of the proposed dark atom scenario.
\end{abstract}

\noindent Keywords: Dark atoms; composite dark matter; stable charged particles; ipole coulomb barrier; effective interaction potential; $X$He; $X$-helium


\section{$X$-helium dark atoms}

The non-baryonic essence of dark matter suggests the existence of new stable forms of non-relativistic matter that play the role of stable dark matter particles in the universe. If dark matter is particle-based, it implies the presence of new stable particles beyond the Standard Model. It has been proposed that such particles may include stable, electrically charged particles \cite{KHLOPOV_2013, Bertone_2005, scott2011searches, belotsky2006composite}. This article discusses the minimal walking technicolor (WTC) model, which introduces a novel perspective on dark matter as a composite entity \cite{Kh_2008, Kh_2013, Beylin2020}. The WTC model suggests the existence of heavy fermions associated with new gauge interactions, and within this framework, the Higgs boson is characterized by composite structure derived from single scalar doublet. In the WTC model, the electric charge of stable, multicharged particles remains undefined, yet stringent experimental constraints dictate that these particles can only possess stable, negatively charged state of $-2n$ \cite{Cudell:2012fw, bulekov2017search}, where $n$ is natural number. We denote these particles as $X$, with the specific case when $X$ has charge of $-2$ is denoted as $O^{--}$. 

This article focuses on composite dark matter scenario in which hypothetical, stable, heavy $X^{-2n}$ particles with lepton-like characteristics (i.e., without QCD interactions or with highly suppressed QCD interactions) form neutral atom-like states with $n$ $^4He$ nuclei of primary helium via usual electromagnetic Coulomb binding. Such configurations are called $X$He dark atoms, where $X^{-2n}$ particles may exhibit lepton-like properties or represent unique combinations of heavy quark new families, marked by weak interactions with hadrons \cite{Khlopov_2020}.

The structural features of bound dark atom system are defined by parameter $a \approx Z_{\alpha} Z_X \alpha A_{\alpha} m_p R_{nHe}$, with $\alpha$ denoting the fine-structure constant, $Z_X$ and $Z_{\alpha}$ are the charge numbers of the $X$ particle and $nHe$ nucleus, $m_{p}$ is the proton mass, $A_{\alpha}$ is the mass number of $nHe$, and $R_{nHe}$ is the radius of the $nHe$ nucleus. Here, $a$ signifies the ratio of the Bohr radius of the dark atom to the radius of the $n$-helium nucleus. When the Bohr radius of the $X$He atom is smaller than the radius of $n$-helium nucleus, the dark atom resembles Thomson-like structure, otherwise, it represents Bohr atom.

For values of $a$ within $0 < a < 1$, the $X$He configuration aligns with Bohr atom model, where the helium nucleus, approximated as point particle, orbits the centrally positioned, negatively charged $X$ particle. Conversely, for $a$ values within $1 < a < \infty$, the structure aligns with Thomson’s atomic model, where the not-point-like helium nucleus oscillates around the heavier negatively charged $X$ particle, reflecting a more distributed atomic configuration. 

The unique characteristics of dark atoms give rise to a "warmer-than-cold dark matter" scenario in the formation of large-scale structures, which, though requiring additional exploration, aligns with data from precision cosmology \cite{Khlopov_2020}. The relevance of this article is expressed by the need for further investigation into the nuclear properties of dark atoms and the potential impacts of $X$-helium on nuclear transformations. Understanding these interactions is crucial for quantifying the role of dark atoms in primary cosmological nucleosynthesis, stellar evolution, and other physical, astrophysical, and cosmological processes in the early universe \cite{Khlopov:2010ik}.

The varied results from direct dark matter detection experiments highlight the complexities in interactions between dark matter particles and materials in underground detectors. The $X$-helium hypothesis suggests that the formation of low-energy bound states between dark atoms and nuclei in detector materials could account for the positive findings of the $DAMA/NaI$ and $DAMA/LIBRA$ experiments, which differ from the negative results observed in $XENON100$, $LUX$, and $CDMS$ \cite{BERNABEI_2020, Khlopov_2020}.

Due to the unscreened nuclear charge of dark atoms, the possibility of strong nuclear interactions between $X$He atoms and matter nuclei could disrupt the bound state of dark atoms, potentially producing anomalous isotopes, whose environmental abundance is highly constrained by experimental limits \cite{Cudell:2012fw}. To address this, the $X$He hypothesis introduces a shallow potential well and a dipole Coulomb barrier within the effective interaction potential between dark atoms and nuclei, which prevents the fusion of $nHe$ and $X$ particles with ordinary matter nuclei. This is crucial condition for the stability  and viability of the $X$-helium hypothesis.

Modeling the interaction between dark atoms and ordinary nuclei presents three-body problem, lacking an exact analytical solution. Thus, to understand the physical implications of this scenario -- defined by a dipole Coulomb barrier and shallow well in the effective interaction potential -- precise quantum mechanical numerical model for this three-body system is being developed. The model aims to reconstruct the effective interaction potential, allowing detailed analysis of the properties and dynamics of the interactions between dark atoms, as composite constituents of dark matter, and nuclei of ordinary matter.

\section{The isolated dark atom system}

It is well understood that dark atom, when exposed to an alternating electric field from external nucleus, experiences the Stark effect, causing polarization of the $X$He atom. This polarization generates dipole Coulomb repulsion between the dark atom and the nucleus, which, in turn, can lead to the establishment of bound state between the $X$-helium and the nucleus of substance. This bound state arises due to potential well that precedes the dipole Coulomb barrier in the total effective interaction potential in $X$He--nucleus system.

To reconstruct the effective interaction potential in the $X$He--nucleus system with accuracy, precise calculation of the Stark potential is crucial. This potential determines the interaction between the polarized dark atom, functioning as $X$He dipole, and the charged heavy nucleus. The Stark potential significantly influences both the depth of the potential well, which defines the low-energy bound state between $X$He and the nucleus of ordinary matter, and the height of the dipole Coulomb barrier, which repels the dark atom from the nucleus and thus prevents their fusion. To accomplish this, quantum mechanical calculations of the dark atom’s dipole moment $\Vec{\delta}$ under the influence of an alternating external electric field (via the Stark effect) are essential, as the Stark potential depends directly on the dipole moment $\Vec{\delta}$ according to the relation:
\begin{equation}
U_{St}=e Z_{nHe} (\Vec{E}_{nuc}\cdot\Vec{\delta}),
\label{eq_Stark}
\end{equation}
where $\Vec{E}_{nuc}$ denotes the strength of the external electric field generated by the heavy charged nucleus of substance, and $Z_{nHe}$ represents the charge number of the $nHe$ nucleus.

In this article, we will examine the special case where the charge of the $X$ particle is -2, such that the $X$ particle is $O^{--}$ particle bound to $^4He$ nucleus of primordial helium, forming neutral $O$He dark atom.

For accurate quantum mechanical calculation of the dipole moment of polarized dark atom, it is necessary not only to obtain the helium wave functions in its ground state within the $O$He--nucleus system but also to determine the ground-state wave function of helium within an isolated, non-polarized $O$He dark atom. Consequently, the initial step involves studying $\hat{H_{0}}$, the Hamiltonian operator of isolated $O$He dark atom, which is free from external influences. Using numerical difference scheme, $\hat{H_{0}}$ is represented as a matrix, and its eigenvalues are computed numerically. These eigenvalues yield the discrete energy levels of helium, $E_{OHe}$, within the isolated $O$He atom, while the eigenvectors corresponding to these states represent the wave functions, $\Psi$, of helium in $O$-helium. This involves solving the following one-dimensional Schrödinger equation:
\begin{equation}
\hat{H_{0}}\Psi(\Vec{r}) = E_{OHe}\Psi(\Vec{r}),
\label{eq_Shred_OHe_isolate}
\end{equation}
or by presenting this expression in another form:
\begin{equation}
\Delta_{r}\Psi(\Vec{r}) + \cfrac{2m_{He}}{\hbar^2}\biggl(E_{OHe} + \cfrac{4e^2}{r}\biggr)\Psi(\Vec{r})=0,
\label{eq_Shred_OHe_isolate_2}
\end{equation}
where $\Vec{r}$ denotes the position vector of the helium nucleus, $m_{He}$ refers to the mass of the helium nucleus, and $\hbar$ is the Planck constant. The coordinate system is centered at the position of the $O^{--}$ particle.

By numerically solving the one-dimensional Schrödinger equation (\ref{eq_Shred_OHe_isolate}) using numerical difference scheme, with the helium radius vector range set to  $r=|2.5~{\times}~10^{-12}~\text{cm}|$ and the number of iterations $N_{iter}=2000$, the first three eigenvalues of the Hamiltonian operator, $\hat{H_{0}}$, were determined: $E_{1,2,3_{num}}= -1.585, -0.393, -0.042 \hspace{2mm}\text{MeV}$. Theoretical calculations for the first three energy levels of helium in the $O$-helium dark atom yield $E_{1,2,3_{OHe}}= -1.589, -0.397, -0.177 \hspace{2mm}\text{MeV}$ \cite{Bikbaev_2024}. As observed, the first two calculated energy levels are consistent with the theoretical values to the second decimal place.
For the purposes of quantum mechanical numerical calculation of the dipole moment of polarized $O$He, it is necessary to know only the wave function corresponding to the first, ground energy level of helium within isolated $O$-helium atom.

\section{Interaction potential of helium in the three-body $O$He-nucleus system}

The three-body problem at hand involves three-body interaction within the $X$He--nucleus system. We focus on the particular case where the $X$He dark atom resembles hydrogen-like Bohr atom, namely $O$-helium. Here, the coordinate system has its origin at the center of the $O^{--}$ particle, which binds to the point-like helium nucleus via Coulomb forces, forming bound atomic system of composite dark matter. The $O$He dark atom is subjected to an inhomogeneous external electric field generated by a third particle, namely, a nucleus characterized by charge number $Z_{nuc}$, neutron number $N_{nuc}$, and mass number $A$. This nucleus approaches the dark atom gradually, engaging in both electromagnetic and strong nuclear interactions.

The Hamiltonian for the point-like helium nucleus in the $O$He -- nucleus system can be expressed as:
\begin{equation}
\hat{H} = \hat{H_{0}} + \hat{U},
\end{equation}
where $\hat{H_{0}}$ represents the Hamiltonian of the isolated $O$He dark atom, unaffected by external forces, and $\hat{U}$ corresponds to the interaction potential between helium and the external nucleus of substance.

We define the vectors $\Vec{r}$, $\Vec{R}_{OA}$, and $\Vec{R}_{HeA}$ as follows: $\Vec{r}$ represents the relative distance vector between the $O^{--}$ particle and the helium nucleus, $\Vec{R}_{OA}$ is the position vector of the external nucleus, and $\Vec{R}_{HeA}$ denotes the vector pointing from the center of the helium nucleus to the center of the external nucleus. These vectors are related by the equation:
\begin{equation}
\Vec{R}_{HeA} = \Vec{R}_{OA} - \Vec{r}.
\end{equation}

Next, let's write down $\hat{H_{0}}$ and $\hat{U}$:
\begin{equation}
\hat{H_{0}} = -\cfrac{\hbar^2}{2m_{He}}\Delta -\cfrac{4e^2}{r},
\end{equation}

\begin{equation}
\hat{U} = U_{Coulomb}(|\Vec{R}_{OA} - \Vec{r}|) +U_{Nuc}(|\Vec{R}_{OA} - \Vec{r}|) + U_{rot_{(He-Na)}}(|\Vec{R}_{OA} - \Vec{r}|),
\end{equation}
here $U_{Nuc}(|\Vec{R}_{OA} - \Vec{r}|)$ signifies the nuclear interaction potential, formulated using the Woods--Saxon potential. $U_{Coulomb} (|\Vec{R}_{OA} - \Vec{r}|)$ describes the Coulomb potential between the point-like helium nucleus and the not-point-like of the external nucleus. The term $U{rot_{(He-Na)}}$ accounts for the centrifugal potential arising from the interaction between helium and sodium nuclei.

The nuclear potential is calculated dependent on the spacing between the neutron distribution surfaces of the interacting nuclei. Specifically, $U_{Nuc}(|\Vec{R}_{OA} - \Vec{r}|)$ is defined by: 
\begin{equation}
    U_{Nuc}(|\Vec{R}_{OA} - \Vec{r}|)=-\cfrac{U_{0}}{1+\exp{\biggl(\cfrac{|\Vec{R}_{OA} - \Vec{r}|-R_{N_{nuc}}-R_{N_{He}}}{p}\biggr)}},
    \label{eq}
\end{equation}
where $R_{N_{nuc}}$ and $R_{N_{He}}$ denote the root-mean-square radii of neutron distributions in the heavy nucleus and helium, respectively, $U_{0}$ is the depth of the potential well (approximately 43 \text{MeV} for sodium), and $p$ is the diffuseness parameter, set to about $0.55 \hspace{2mm} \text{fm}$.

The radii $R_{N_{nuc}}$ and $R_{N_{He}}$ are computed using the following expressions \cite{Seif_2015}:
\begin{equation}
    R_{N_{nuc,He}}=\sqrt{\cfrac{3}{5}R_{0N_{nuc,He}}^2 + \cfrac{7\pi^2}{5}a_{N_{nuc,He}}^2}\sqrt{1 + \cfrac{5b_{nuc,He}^2}{4\pi}} \hspace{2mm}\text{fm},
    \label{eq}
\end{equation}
where $b_{nuc,He}$ denotes the deformation parameter for both the heavy nucleus of the substance and the helium nucleus. For the sodium nucleus, this deformation parameter is assigned value of $b_{Na}=0.447$, while the helium nucleus is considered spherically symmetric, giving it deformation parameter of zero.
The variable $R_{0N_{nuc,He}}$ represents the half-radius of the neutron distribution for both the heavy nucleus of matter and the helium nucleus. This radius is calculated based on the neutron number $N$ and proton number $Z$ of the respective nucleus using the formula: 
\begin{equation}
R_{0N_{nuc,He}}=0.953N_{nuc,He}^{1/3}+0.015Z_{nuc,He}+0.774 \hspace{2mm}\text{fm},
    \label{eq}
\end{equation}
and the parameter $a_{N_{nuc,He}}$ is dimensional constant related to the proton and neutron counts $Z$ and $N$ of respective nucleus, and it is determined according to the following expression:
\begin{equation}
    a_{N_{nuc,He}}=0.446 + 0.072\cfrac{N_{nuc,He}}{Z_{nuc,He}} \hspace{2mm}\text{fm}.
    \label{eq}
\end{equation} 

The Coulomb interaction potential, $U_{Coulomb}(|\Vec{R}_{OA} - \Vec{r}|)$, between the point-like helium nucleus and the nucleus of heavy element, whose radius corresponds to the root-mean-square radius of proton distribution $R_{p_{nuc}}$, is given by:
\begin{equation}
U_{Coulomb}(|\Vec{R}_{OA} - \Vec{r}|)=
\begin{cases}
\cfrac{2e^2Z_{nuc}}{|\Vec{R}_{OA} - \Vec{r}|} & \text{for}\hspace{3pt} |\Vec{R}_{OA} - \Vec{r}|>R_{p_{nuc}},\\
\cfrac{2e^2Z_{nuc}}{2R_{p_{nuc}}}\left(3-\cfrac{|\Vec{R}_{OA} - \Vec{r}|^2}{R_{p_{nuc}}^2}\right) & \text{for}\hspace{3pt} |\Vec{R}_{OA} - \Vec{r}|<=R_{p_{nuc}},
\end{cases}
\end{equation}
where the radius $R_{p_{nuc}}$ is defined according to the expression \cite{Seif_2015}:
\begin{equation}
    R_{p_{nuc}}=\sqrt{\cfrac{3}{5}R_{0p_{nuc}}^2 + \cfrac{7\pi^2}{5}a_{p_{nuc}}^2}\sqrt{1 + \cfrac{5b_{nuc}^2}{4\pi}} \hspace{2mm}\text{fm},
    \label{eq}
\end{equation}
here, the parameter $R_{0p_{nuc}}$ denotes the half-radius of the proton distribution within the nucleus and is calculated as function of the charge number $Z_{nuc}$ and neutron number $N_{nuc}$ for the heavy nucleus as follows:
\begin{equation}
R_{0p_{nuc}}=1.322Z_{nuc}^{1/3}+0.007N_{nuc}+0.022 \hspace{2mm}\text{fm},
    \label{eq}
\end{equation}
the constant $a_{p_{nuc}}$ is dimensional parameter that also depends on the number of protons and neutrons of the nucleus, expressed by the relation:
\begin{equation}
    a_{p_{nuc}}=0.449 + 0.071\cfrac{Z_{nuc}}{N_{nuc}} \hspace{2mm}\text{fm}.
    \label{eq}
\end{equation} 

$U_{rot_{(He-Na)}}(|\Vec{R}_{OA} - \Vec{r}|)$ is calculated by the formula (refer to formula 27 in \cite{Adamian_1996}):
\begin{equation}
    U_{rot_{(He-Na)}}(|\Vec{R}_{OA} - \Vec{r}|)=\cfrac{\hbar^2c^2J_{(He-Na)}(J_{(He-Na)}+1)}{2 m_{He} c^2 |\Vec{R}_{OA} - \Vec{r}|^2},
    \label{eq}
\end{equation}
in this context, $\Vec{J}_{(He-Na)}$ represents the total angular momentum of the helium and sodium nuclei in their interaction.

The total angular momentum of the helium-sodium interaction, $\Vec{J}_{(He-Na)}$, is equal to the intrinsic angular momentum of the sodium nucleus, $\Vec{I}_{Na}=\overrightarrow{3/2}$. Since the helium nucleus has intrinsic angular momentum of $\Vec {I}_{He}=\Vec{0}$ and because the impact parameter of the sodium nucleus approaching the helium nucleus is zero, the orbital angular momentum between the helium and sodium nuclei is also zero. Thus, we obtain $\Vec{J}_{(He-Na)} = \overrightarrow{3/2}$.

Accordingly, the Hamiltonian $\hat{H}$ for the helium nucleus within the $O$He--nucleus system is determined by the radius vectors $\Vec{r}$ and $\Vec{R}_{OA}$. However, by setting $\Vec{R}_{OA}$ as fixed value and incrementally changing the external nucleus position (i.e., by varying $\Vec{R}_{OA}$), set of Schrödinger equations dependent on $\Vec{r}$ can be derived, each equation corresponding to certain set position of the external nucleus relative to the dark atom.

Thus, the Schrödinger equation to be solved takes the form:
\begin{equation}
\hat{H}\Psi(\Vec{r}) = E\Psi(\Vec{r}),
\end{equation}
which, upon expanding $\hat{H}$ and applying relevant transformations, results in the following expression:
\begin{equation}
\begin{split}
\Delta\Psi(\Vec{r}) + \cfrac{2m_{He}}{\hbar^2}\biggl(E + \cfrac{4e^2}{r} &- U_{Coulomb}(|\Vec{R}_{OA} - \Vec{r}|) - U_{N}(|\Vec{R}_{OA} - \Vec{r}|) - \\ & -U_{rot_{(He-Na)}}(|\Vec{R}_{OA} -\Vec{r}|)\biggr)\Psi(\Vec{r})=0.
\end{split}
\label{eq_Shred_OHe_nuc}
\end{equation}

To numerically determine the eigenvalues of the Hamiltonian operator $\hat{H}$, which correspond to the energy levels of helium $E$ within the $O$He--nucleus system for each fixed position $\Vec{R}_{OA}$ of the external nucleus, we approximate $\hat{H}$ using finite difference operator in matrix form. This approach also allows the calculation of the Hamiltonian's eigenvectors, which represent the helium wave functions $\Psi$ for this system.

To achieve this, in addition to expressing the Laplace operator in matrix form, we must construct the matrix representation of the interaction potential of the helium nucleus within the $O$He--nucleus system for each specific fixed  value of $\Vec{R}_{OA}$:
\begin{equation}
U_{He} = -\cfrac{4e^2}{r} + U_{Coulomb}(|\Vec{R}_{OA} - \Vec{r}|) + U_{N}(|\Vec{R}_{OA} - \Vec{r}|) + U_{rot_{(He-Na)}}(|\Vec{R}_{OA} - \Vec{r}|).
\label{eq_U_sum_He}
\end{equation}

\begin{figure}[h!]
\centering
\includegraphics[scale=0.37]{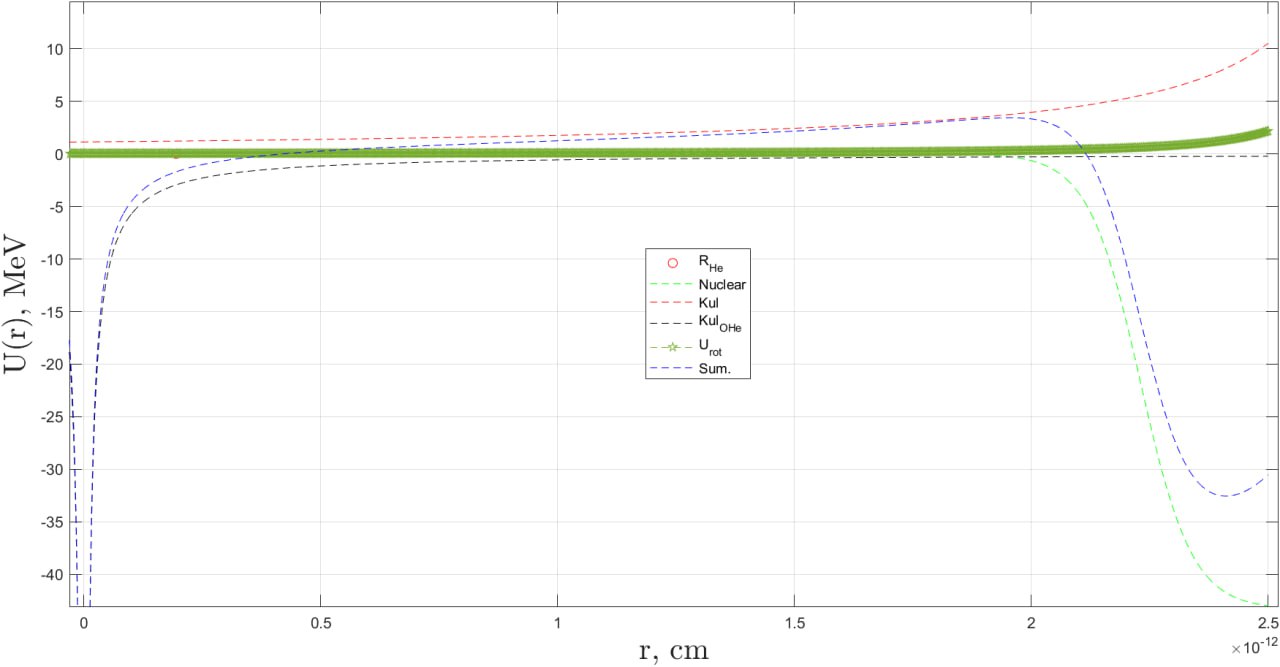}
\caption{Potentials of Coulomb (red dotted line), nuclear (green dotted line) and centrifugal (green solid line) interaction between helium and the nucleus of $Na$, the potential of Coulomb interaction between helium and $O^{--}$ particle (black dotted line) and the total interaction potential of the helium nucleus (blue dotted line) in the $O$He--$Na$ system at fixed $\Vec{R}_{OA}$. The red circle marks the value of the radius of the $He$ nucleus. Original authors’ figure taken from \cite{Bikbaev_2024}.}
\label{fig:Ris27}
\end{figure}

Figure \ref{fig:Ris27} presents example of the reconstructed total interaction potential, $U_{He}$, for helium within the $O$He--$Na$ system as function of the helium radius vector $\Vec{r}$, while keeping the radius vector $\Vec{R}_{OA}$ of the external sodium nucleus fixed.

Figure \ref{fig:Ris27} illustrates the Coulomb and nuclear interaction potentials between the helium and sodium nuclei, also the centrifugal potential for helium--sodium interaction at zero impact parameter, $U_{rot_{(He-Na)}}(|\Vec{R}_{OA} - \Vec{r}|)$. Additionally, it shows the Coulomb interaction potential between helium and the $O^{--}$ particle, alongside the total interaction potential for helium in the $O$He--$Na$ system.

Thus, the quantum mechanical numerical approach to solving the three-body problem in the $O$He--nucleus system involves resolving the Schrödinger equation for the helium nucleus within the $O$He--nucleus framework for each fixed external nucleus position, $\Vec{R}_{OA}$. This requires expressing the Hamiltonian of the helium nucleus in matrix form and performing numerical calculations of its eigenvalues and eigenvectors, which represent the energy levels and wave functions ($\Psi$-functions) of helium in the $O$He--nucleus system, respectively.

\section{Calculation of the dipole moment values of polarized dark atom}

When external nucleus is not present, the dark matter atom remains unpolarized, with the helium energy level in the ground state of $O$He around $1.6 \hspace{0.2 cm}\text{MeV}$. However, as an external nucleus approaches, the varying electric field from this heavy nucleus induces the Stark effect, causing polarization of the dark atom. Consequently, $O$He develops non-zero dipole moment and begins to interact with the external nucleus as dipole. This interaction can be described by the Stark potential, as shown in Equation \eqref{eq_Stark}. According to the dark atom model, dipole barrier is expected to form within the effective interaction potential between $O$He and the heavy nucleus of substance, preventing the fusion of dark matter particles with the nucleus. Furthermore, low-energy bound state between the dark atom and the heavy nucleus should also emerge.

In addressing the one-dimensional Schrödinger equation (SE) for the helium nucleus in the $O$He--nucleus system (see Equation \eqref{eq_Shred_OHe_nuc}), it is essential to define the range for the helium radius vector $\Vec{r}$ with the external nucleus position $\Vec{R}_{OA}$ held constant. Here, $\Vec{r}$ acts as free parameter that determines the shape of the total interaction potential for helium in the $O$He--nucleus system, in which the corresponding SE to be solved for each specified fixed location of the heavy nucleus. To solve this set of SE -- SE for each fixed position of the slowly approaching external nucleus -- the interval for the radius vector of the heavy nucleus, $\Vec{R}_{OA}$, must also be determined.

By overlapping $\Vec{r}$ with $\Vec{R}_{OA}$, helium would likely reside within the deep potential well created by the heavy nucleus. Given that the helium nucleus is situated within the dark atom -- where $O$He forms bound quantum mechanical system prior to its interaction with the heavy nucleus begins -- the ranges for $\Vec{r}$ and $\Vec{R}_{OA}$ should be chosen that their boundaries to be close in proximity without overlapping. This configuration ensures that the helium nucleus remains part of the dark atom initially, gradually sensing the influence of the approaching nucleus. As the heavy nucleus draws nearer, the probability of the helium nucleus tunneling through the Coulomb barrier into the nucleus increases. 
Therefore, for the specified interval of $\Vec{r}$, defined by boundary points that are equal in magnitude yet opposite in sign, the interval for $\Vec{R}_{OA}$ is set to begin at considerable distance from the dark atom and to end close to the right endpoint of the helium radius vector interval. In this case, the radius vector for helium is represented as $\Vec{r}=[-a; a]$, while the radius vector for the external nucleus is expressed as $\Vec{R}_{OA}=[c; b]$, where $a, c, b > 0$ and $a \leqslant b < c$. Consequently, the fixed position of the external nucleus, $\Vec{R}_{OA}$, will consistently take values within the interval $[c; b]$, progressing from point $c$ to point $b$. For each point $p^{*}\in[c; b]$, the distance between the helium nucleus and the matter nucleus, $\Vec{R}_{HeA} = \Vec{R}_{OA} - \Vec{r}$, will vary within the interval $\Vec{R}_{HeA} = [p^{*}+a; p^{*}-a]$. As the external nucleus approaches the dark matter atom, the polarization of $O$He is expected to increase in response to the nucleus's proximity.

As the nucleus of the substance moves closer to dark atom, the ground state of the helium nucleus within $O$-helium undergoes corresponding shifts. To calculate changes in the dipole moment of the polarized dark atom, it is necessary to calculate the shifts in the energy of the ground state and the corresponding wave functions of the polarized dark atom.

By solving the set of Schrödinger equations for helium in the $O$He--$Na$ system for various fixed positions of the heavy nucleus of matter $\Vec{R}_{OA}$, we have obtained set of energy values of the ground state of helium corresponding to certain polarization of the $O$He atom at certain fixed position of the outer nucleus of matter and the wave functions of helium corresponding to these ground states of the polarized dark atom.

Utilizing the normalized ground-state wave function of helium in unpolarized dark atom, $\Psi_{OHe}$, obtained by solving the Schrödinger equation for helium in the isolated $O$-helium dark atom, along with the normalized wave functions of helium in the polarized dark atom for different ground-state energy values, $\Psi_{OHeNa}$, we determined the spectrum of dipole moment values $\delta$ for the polarized $O$He. The dipole moment $\delta$ corresponding to each $\Psi_{OHeNa}$ was computed as follows:
\begin{equation}
    \delta= \int_r \Psi_{OHe}^{*}\cdot r \cdot \Psi_{OHeNa} \cdot 4\pi r^{2} dr.
    \label{eq:delta}
\end{equation}

To accurately evaluate the integral in (\ref{eq:delta}), precise determination of its integration limits is required. Since we are calculating the dipole moments of the polarized dark atom, it is crucial to account for the probability distribution of locating the helium nucleus within the dark atom. To define the left and right bounds of integration, we must identify the intersection points between the plot of the squared modulus of the helium wave function and the plot of the total helium potential within the $O$He -- $Na$ system at fixed value of $\Vec{R}_{OA}$. For each fixed position of the external heavy nucleus, this approach allows us to set the integration region within the dark atom, effectively establishing the integration limits. These limits are defined by the intersection points where the graphs of the total helium interaction potential and the squared modulus of the wave function, associated with specific ground-state energy level, meet.

\begin{figure}[h]
\centering
\includegraphics[scale=0.37]{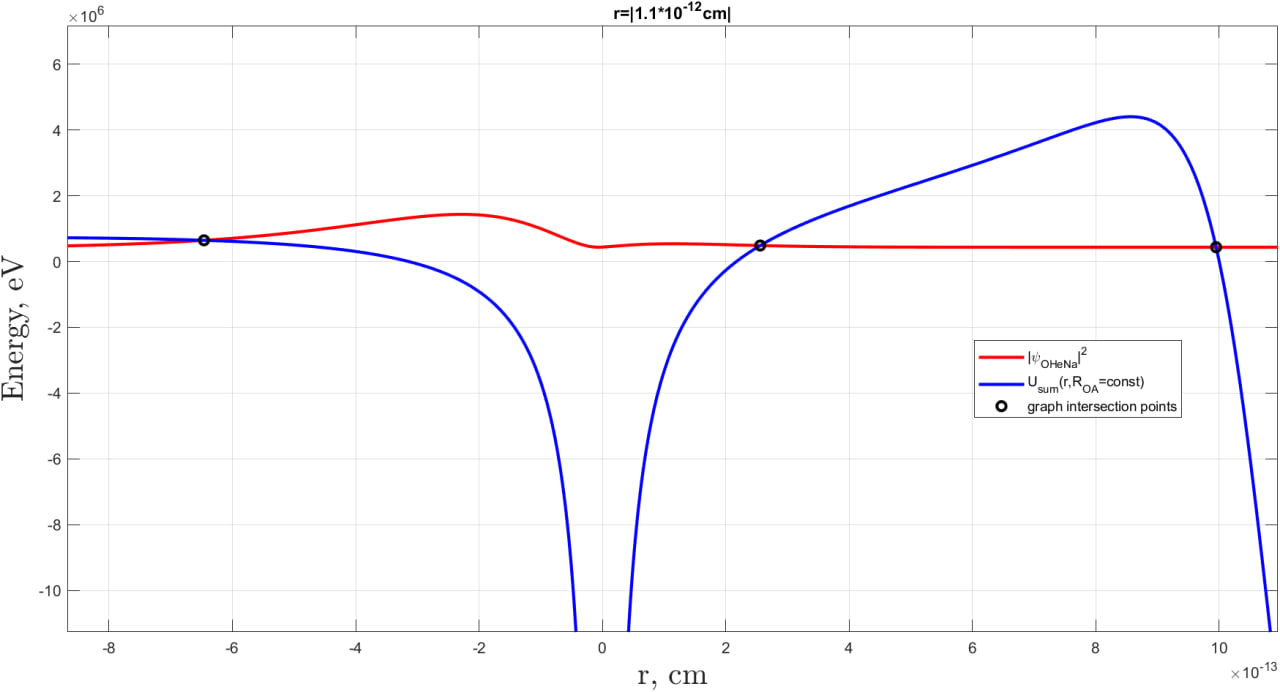}
\caption{The total potential of helium in the $O$He--$Na$ system for fixed position of sodium $\Vec{R}_{OA}$ (blue solid line), graph of the squared modulus of the wave function of the ground state of helium in polarized dark atom for fixed $\Vec{R}_{OA}$ (red solid line), the intersection points of the graph of the total potential of helium and the graph of the squared modulus of the wave function of the ground state of helium (black circles). Original authors’ figure taken from \cite{Bikbaev_2024}.}
\label{fig:Ris19}
\end{figure}

Figure \ref{fig:Ris19} illustrates the method for determining these integration limits necessary for calculating the integral in (\ref{eq:delta}). In the figure, the blue solid line represents the total interaction potential of helium within the $O$He -- $Na$ system for specific fixed position of the sodium nucleus, $\Vec{R}_{OA}$, while the red solid line shows the squared modulus of the helium ground-state wave function within the polarized dark atom at this fixed $\Vec{R}_{OA}$. The black circles mark the intersection points of the two curves, with the first two intersections from left to right indicating the integration bounds for (\ref{eq:delta}). In the example shown in Figure \ref{fig:Ris19}, the dark atom is negatively polarized, as the probability density of locating helium to the left of the origin (or the $O^{--}$ particle) is greater than that on the right. Here, the sodium nucleus is positioned close enough that the helium begins to experience the nuclear potential (visible as potential well forming to the right of the Coulomb barrier), but not so close as to result in significant tunneling of helium through the barrier.

\unskip

\begin{figure}[h]
\centering
\includegraphics[scale=0.37]{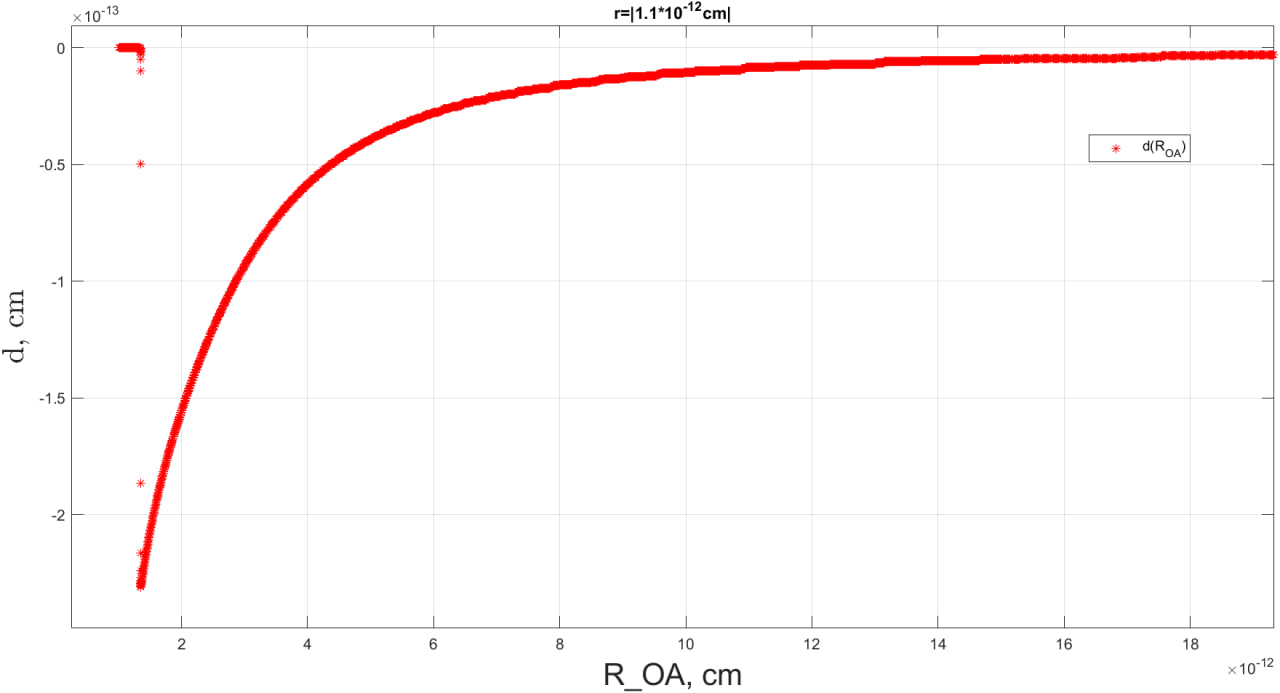}
\caption{Graph of the dependence of the dipole moment of polarized $O$He atom (red stars) on the radius vector of the outer sodium nucleus. Original authors’ figure taken from \cite{Bikbaev_2024}.}
\label{fig:Ris12}
\end{figure}

By calculating the spectrum of dipole moment values, $\delta$, for polarized $O$He at various positions of the sodium nucleus $\Vec{R}_{OA}$, we can illustrate how the dipole moment of the polarized dark atom varies with the radius vector $\Vec{R}_{OA}$ (as depicted in Figure \ref{fig:Ris12}).

In Figure \ref{fig:Ris12}, red stars indicate the values of the dipole moment for the polarized $O$He atom, each corresponding to specific fixed values of $\Vec{R}_{OA}$ within the helium radius vector interval $r=|1.1\times10^{-12} ~\text{cm}|$. From Figure \ref{fig:Ris12}, it can be observed that when the sodium nucleus is distant from $O$He, the dark atom behaves as isolated system, with dipole moment approaching zero. As the sodium nucleus moves closer to the $O$-helium atom, $O$He becomes more polarized, resulting in progressively larger negative dipole moment. This increase in polarization occurs because the sodium nucleus exerts repulsive Coulomb force on helium, encouraging the helium to shift leftward relative to the $O^{--}$ particle. When the sodium nucleus is in close proximity to the dark atom and $\Vec{R}_{OA}$ nears the right boundary of the helium radius vector interval $\Vec{r}$, the nuclear force between helium and sodium dominates over the Coulomb interaction, causing $\delta$ to approach above zero.

\section{Reconstruction the total effective interaction potential of the $O$He -- $Na$ system}

To reconstruct the Stark potential form, as given in Formula \eqref{eq_Stark}, which describes the electric interaction between the dipole of polarized dark atom and substance nucleus, we employ the dipole moment values calculated through quantum mechanical methods. Additionally, to obtain the total effective interaction potential for the $O$He--$Na$ system (the total interaction potential between the sodium nucleus and the dark atom of $O$-helium), it is necessary also to reconstruct the form of the nuclear interaction potential, this is achieved by using the Woods–Saxon model for helium and sodium nuclei, along with the electric interaction potential, $U_{XHe}^{e}$, between the unpolarized $O$He dark atom and the sodium nucleus. The latter potential, $U_{XHe}^{e}$, is derived by solving the self-consistent Poisson equation, accounting for the screening effect of the $O^{--}$ particle by the helium nucleus. This screening effect becomes significant only at close distances to the dark atom, as it decays exponentially (see Section 5.1 on the "Approach of Reconstructing of Interaction Potentials in the $X$He--nucleus system" in \cite{Bikbaev_2023}).
In addition, in order to reconstruct the shape of the total effective interaction potential of the sodium nucleus with the dark atom, we must also take into account the potential of the centrifugal interaction of the sodium nucleus with the dark atom of $O$-helium.

The centrifugal potential of the interaction of the $O$He dark atom with the sodium nucleus, denoted as $U_{rot_{(OHe-Na)}}$, is determined by the total angular momentum of the system of  interacting particles, $\Vec{J}_{(OHe-Na)}$, as well as the distance between the interacting particles, $R$. Ignoring the moments of inertia of the nuclei, it is defined by the following expression (see formula 27 in \cite{Adamian_1996}):
\begin{equation}
    U_{rot_{(OHe-Na)}}(R)=\cfrac{\hbar^2c^2J_{(OHe-Na)}(J_{(OHe-Na)}+1)}{2 \mu c^2 R^2},
    \label{eq}
\end{equation}
where $\mu$ represents the reduced mass of the interacting particles.

Since the mass of $O$He is entirely determined by the mass of the heavy particle $O^{--}$, taken as $1\hspace{2mm} \text{TeV}$, and the sodium nucleus has mass of approximately  $m_{Na} \approx 21.4\hspace{2mm} \text{GeV}$, which is significantly smaller in comparison with $O$He’s mass, the reduced mass of the system can be approximated by the sodium mass, $\mu \approx m_{Na}/c^2$.

The total angular momentum of the interacting particles, $\Vec{J}_{(OHe-Na)}$, is given by:
\begin{equation}
    \Vec{J}_{(OHe-Na)}(\rho)=\Vec{l}_{(OHe-Na)}(\rho) + \Vec{I}_{Na} + \Vec{I}_{OHe},
    \label{eq}
\end{equation}
here $\Vec{l}_{(OHe-Na)}(\rho)$ denotes the orbital angular momentum, which depends on the impact parameter $\rho$, $\Vec{I}_{Na}$ represents the intrinsic angular momentum of the sodium nucleus, and $\Vec{I}_{OHe}$ signifies the spin of the $O$He dark atom. The spin $\Vec{I}_{OHe}$ is defined as the sum of vectors of the spin of the $O^{--}$ particle, $\Vec{I}_{O^{--}}$, and the intrinsic angular momentum of the helium nucleus, $\Vec{I}_{He}$.

We consider the case of frontal collision between the sodium nucleus and the $O$He dark atom. In this head-on collision, where the sodium nucleus approaches with zero impact parameter, $\rho = 0$, the orbital angular momentum of the interacting particles also  becomes zero, $\Vec{l}_{(OHe-Na)}(0)=\Vec{0}$. The intrinsic angular momentum of the sodium nucleus is $\Vec{I}_{Na} = \overrightarrow{3/2}$.

Thus, for the scenario with an impact parameter $\rho = 0$, the total angular momentum of system of the $O$He dark atom and sodium nucleus, $\Vec{J}_{(OHe -Na)}$, is expressed as:
\begin{equation}
    \Vec{J}_{(OHe-Na)}= \cfrac{\vec3}{2}  + \Vec{I}_{O^{--}}.
    \label{eq_OHe_Na_J}
\end{equation}

If $O^{--}$ is technibaryon, its spin, $\Vec{I}_{O^{--}}$, could be either $\Vec{0}$ or $\Vec{1}$, and if $O^{--}$ belonged to technilepton particle, its spin would be $\Vec{I}_{O^{--}} = \overrightarrow{1/2}$ \cite{WTC_Sannino_2005, Kouv3}. if $O^{--}$ takes the form of $\Delta_{\bar{U}\bar{U}\bar{U}}^{--}$ and includes new quarks from extended families, then $\Vec{I}_{O^{--}} = \overrightarrow{3/2}$ \cite{Belotsky2008}.

\begin{figure}[h!]
\centering
\includegraphics[scale=0.21]{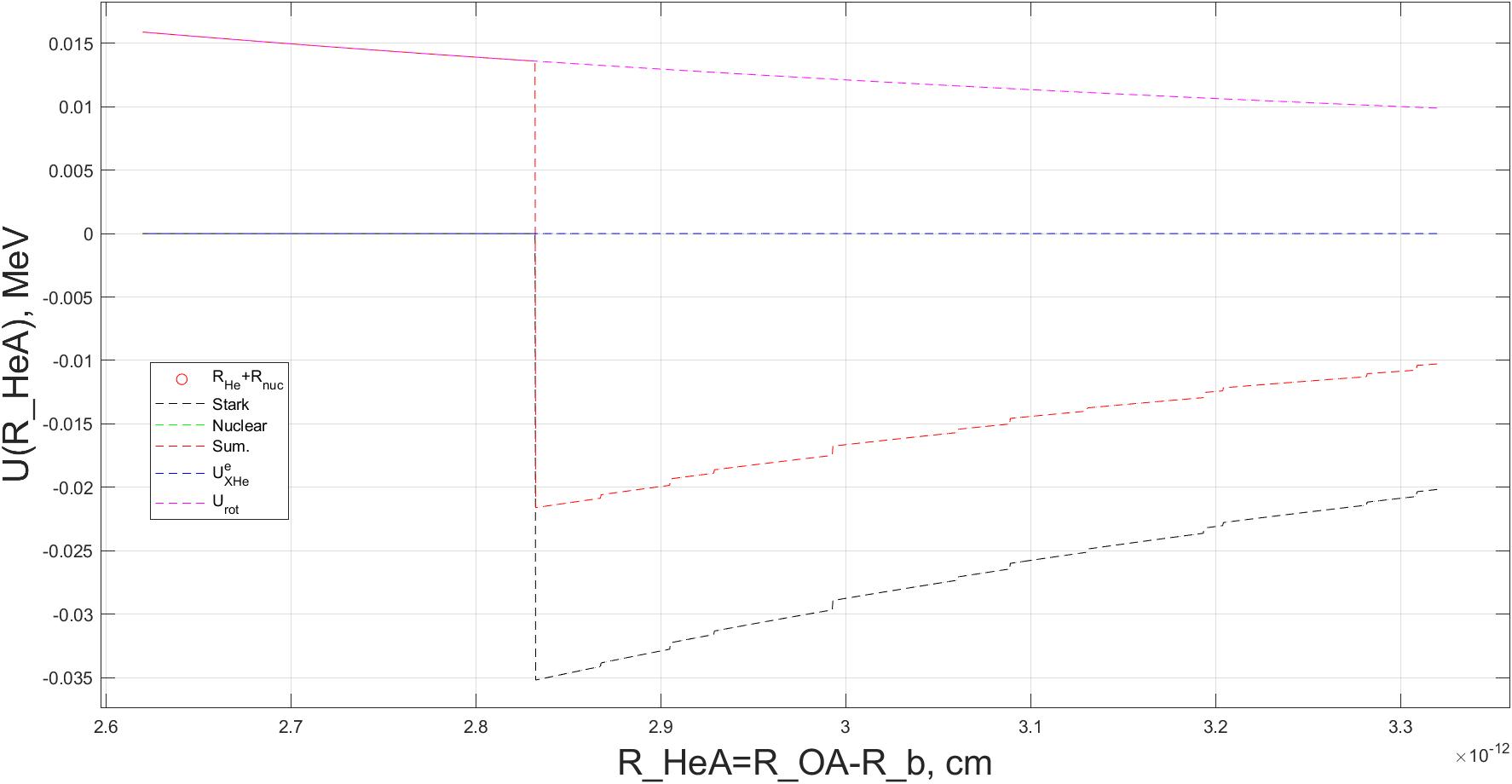}
\caption{
Graphs of Woods--Saxon nuclear potential (green dotted line), $U^{e}_{XHe}$ (blue dotted line), Stark potential (gray dotted line), centrifugal potential (purple dotted line) and total effective interaction potential of $O$He with the nucleus of the sodium (red dotted line) on the distance between the $He$ nucleus, located in the Bohr orbit of the $O$He atom, and the $Na$ nucleus for $J_{(OHe-Na)} = 3$. 
Original authors’ figure taken from \cite{Bikbaev_2024}.
}
\label{fig:Ris29}
\end{figure}

As a result, summing up the potential of the Woods–Saxon nuclear interaction, $U_{XHe}^{e}$, $U_{St}$ and $U_{rot_{(OHe-Na)}}$, we obtain the total effective interaction potential of the $O$-helium dark atom with the sodium nucleus (see Figure~\ref{fig:Ris29}).

The $O$He model suggests that the magnitude of the dipole Coulomb barrier in the total effective interaction potential of the $O$He--$Na$ system should be sufficient to prevent direct fusion between the dark atom and the sodium nucleus. Under the conditions of the $DAMA$ experiment, the relative velocity of the sodium nucleus in the $O$He--$Na$ system is thermal, corresponding to normal room temperature (around $300 \hspace{2mm}\text{K}$). This thermal motion corresponds to kinetic energy of roughly $\sim 2.6\times10^{-2} \hspace{2mm}\text{eV}$ for the sodium nucleus. As result, the height of dipole Coulomb barrier in the effective interaction potential of the $O$He--$Na$ system is expected to be higher than this kinetic energy of sodium nucleus.

Finally, we can construct the total effective interaction potential of $O$He with the sodium nucleus in the $O$He--$Na$ system, for example, for the value of the total angular momentum of the system of interacting particles $\Vec{J}_{(OHe -Na)}=\overrightarrow{3}$, which corresponds to the value of the spin of the $O^{--}$ particle, $\Vec{I}_{O^{--}} = \overrightarrow{3/2}$, as shown in Figure \ref{fig:Ris29}.
In general case, the shape of this total effective interaction potential strongly depends on the value of the spin of the $O^{--}$ particle, $I_{O^{--}}$, however, in all cases, the shape of the total effective interaction potential between sodium and $O$He qualitatively corresponds to theoretical expectations.
This allows for extended range of the helium radius vector interval, $\Vec{r}$, to get the depth of potential well near $\sim 6 \hspace{2mm}\text{keV}$ and positive potential barrier height that exceeds zero as well as greater than the thermal kinetic energy of sodium $\sim 2.6\times10^{-2} \hspace{2mm}\text{eV}$. This barrier height and depth of potential well align with theoretical predictions and experimental data. Such positive potential barrier value playing critical role in preventing the fusion of $He$ and/or $O^{--}$ particles with atomic nuclei, thereby preserving the stability of the dark atom. 
The presence of this barrier in the total effective interaction potential of the $O$He--nucleus system, which ensures the impossibility of direct fusion of dark atom with ordinary atomic nuclei of matter, is fundamental requirement for sustaining the viability of the $O$He dark atom model.

\section{Conclusions}

The numerical model presented in this article is founded on quantum mechanical numerical approximation to describe three-particle system, where interactions occur through electric, centrifugal, and nuclear forces. This method entails solving the Schrödinger equation for helium in the $O$He--$Na$ system for each fixed position of the sodium nucleus relative to the $O$-helium atom. By considering the distinctive features of nuclear and electromagnetic interactions within $O$He--$Na$ system, we can precisely determine the polarization of dark matter atom, calculate the dipole moments of polarized $O$He atom as function of the distance between the dark atom and the sodium nucleus, and, thereby, accurately reconstruct the Stark potential, which plays crucial role in shaping the effective interaction potential in the $O$He--$Na$ system.

Thus, within the framework of the developed numerical model, the helium interaction potential in the $O$He--$Na$ system has been reconstructed, allowing for the solution of the Schrödinger equation for the helium nucleus in both isolated, unpolarized dark atom and polarized $O$-helium within the $O$He--$Na$ system. Then, dipole moments of the polarized $O$He atom were derived using the helium wave functions for both isolated dark matter atom and polarized dark atom in the $O$He--$Na$ system. After that, the Stark potential was computed, enabling the reconstruction of the total effective interaction potential between the sodium nucleus and the $O$He dark atom in the $O$He--$Na$ system. This total effective potential includes contributions from the Stark potential, nuclear potential, centrifugal potential, and the electric interaction potential of unpolarized $O$He dark atom with the sodium nucleus and this total effective potential qualitatively coincides with its theoretically expected form.

To improve the precision of the effective interaction potential reconstruction and achieve more accurate physical model of dark atom interaction with heavy nucleus, as well as to explain the findings of direct dark matter detection experiments, refinements to the quantum mechanical approach to the total effective interaction potential reconstruction are planned. These improvements will involve determining the nuclear and electromagnetic potentials for interaction between $X$-helium and ordinary matter nucleus, with accounting for the finite sizes of the interacting particles by incorporating electric charge and nucleon distributions within the nuclei. Additionally, the model will incorporate nuclei deformation by modeling the nuclei as spherically asymmetric.

\section*{Acknowledgements}
The work by A.M. was performed with the financial support provided by the Russian Ministry of Science and Higher Education, project “Fundamental and applied research of cosmic rays”, No.~FSWU-2023-0068. 




\begin{thebibliography}{99}

\bibitem{KHLOPOV_2013} Khlopov, M. Fundamental particle structure in the cosmological dark matter. {\it International Journal of Modern Physics A} {\bf 2013}, {\it 28}, 1330042. 

\bibitem{Bertone_2005} Bertone, G.; Hooper, D.; Silk, J. Particle dark matter: evidence, candidates and constraints. {\it Physics Reports} {\bf 2005}, {\it 405}, 279 ~-- 390.

\bibitem{scott2011searches} Scott, P. Searches for Particle Dark Matter: An Introduction. {\it arXiv} {\bf 2011}, arXiv:1110.2757.   

\bibitem{belotsky2006composite} Belotsky, K. M.; Khlopov, M.Y.; Shibaev, K. I. Composite Dark Matter and its Charged Constituents. {\it Grav.Cosmol.} {\bf 2006}, {\it 12}, 93-99, arXiv:astro-ph/0604518.

\bibitem{Kh_2008} Khlopov, M. Yu.; Kouvaris, C. Composite dark matter from a model with composite Higgsboson. {\it Phys. Rev.} {\bf 2008}, {\it 78}, 065040.

\bibitem{Kh_2013} Fargion, D.; Khlopov, M. Yu. Tera-leptons’ shadows over Sinister Universe. {\it Gravitation Cosmol.} {\bf 2013}, {\it 19}, 219.

\bibitem{Beylin2020} Beylin, V.; Khlopov, M.; Kuksa, V.; Volchanskiy, N. New Physics of Strong Interaction and Dark Universe. {\it Universe} {\bf 2020}, {\it 6}, 196.

\bibitem{Cudell:2012fw} Cudell, J. R.; Khlopov, M. Y.; Wallemacq, Q. The nuclear physics of OHe. {\it Bled Workshops in Physics} {\bf 2012}, {\it 13}, 10~--27.

\bibitem{bulekov2017search} Bulekov, O. V.; Khlopov, M. Yu.; Romaniouk, A. S.; Smirnov, Yu. S. Search for Double Charged Particles as Direct Test for Dark Atom Constituents. {\it Bled Workshops in Physics} {\bf 2017}, {\it 18}, 11-24.

\bibitem{Khlopov_2020} Khlopov, M. What comes after the Standard Model? {\it Prog. Part. Nucl. Phys.} {\bf 2020}, {\it 116}, 103824.

\bibitem{Khlopov:2010ik} Khlopov, M.Y.; Mayorov, A.G.; Soldatov, E.Y. The dark atoms of dark matter. {\it Prespace J.} {\bf 2010}, {\it 1}, 1403--1417.

\bibitem{BERNABEI_2020} Bernabei, R.;  Belli, P.; Bussolotti, A.; Cappella, F.; Caracciolo, V.; Cerulli, R.; Dai, C.J.; d’Angelo, A.; Di Marco, A.; Ferrari, N.; et al.  The DAMA project: Achievements, implications and perspectives. {\it Prog. Part. Nucl. Phys.} {\bf 2020}, {\it 114}, 103810.

\bibitem{Bikbaev_2024} Beylin, V. A.; Bikbaev, T. E.; Khlopov, M. Y.; Mayorov, A. G.; Sopin, D. O. Dark Atoms of Nuclear Interacting Dark Matter. {\it Universe} {\bf 2024}, {\it 10}, 368.

\bibitem{Seif_2015} Seif, W. Mansour, Hesham. Systematics of nucleon density distributions and neutron skin of nuclei. {\it Int. J. Mod. Phys. E} {\bf 2015}, {\it 24}, 1550083.

\bibitem{Adamian_1996} Adamian, G.G.; Antonenko, N.V.; Jolos, R.V.; Ivanova, S.P.; Melnikova, O.I. Effective nucleus-nucleus potential for calculation of potential energy of a dinuclear system. {\it Int. J. Mod. Phys.  E} {\bf 1996} {\it 5},  191--216. 

\bibitem{Bikbaev_2023} Bikbaev, T.; Khlopov, M.; Mayorov, A. Numerical Modeling of the Interaction of Dark Atoms with Nuclei to Solve the Problem of Direct Dark Matter Search. {\it Symmetry} {\bf 2023}, {\it 15}, 2182.

\bibitem{WTC_Sannino_2005}  Sannino, F.; Tuominen, K.; Orienfold Theory Dynamics and Symmetry Breaking. {\it Phys. Rev. D} {\bf 2005}, {\it 71}, 051901. 

\bibitem{Kouv3} Khlopov, M.Y.; Kouvaris, C. Strong interactive massive particles from a strong coupled theory. {\it Phys. Rev. D} {\bf 2008}, {\it 77}, 065002.

\bibitem{Belotsky2008} Belotsky, K.; Khlopov, M.; Shibaev, K. Stable quarks of the 4th family? {\it arXiv} {\bf 2008}, arXiv:0806.1067.





















\end{thebibliography}
\end{document}